\documentclass[%
 reprint,
 amsmath,amssymb,
 aps,
prr,
superscriptaddress]{revtex4-2}

\newcommand{\ddroit}{\mathrm{d}}
\usepackage{bm}
\usepackage{dcolumn}
\usepackage{graphicx}
\usepackage{xcolor}

\begin{document}

\title{Unveiling the Dynamics of Optical Frequency Combs from Phase-Amplitude Correlations}

\author{M. Ansquer}
 \affiliation{Laboratoire Kastler Brossel, Sorbonne Université, ENS-Université PSL, CNRS, Collège de France, 4 place Jussieu, 75252 Paris, France}
 
\author{V. Thiel}
 \affiliation{Department of Physics and Oregon Center for Optical, Molecular, and Quantum Science, University of Oregon, Eugene, Oregon 97403, USA}
 
\author{S. De}
 \affiliation{Integrated Quantum Optics Group, Applied Physics, 100 Warburger Stra{\ss}e, Paderborn University, Paderborn 33098, Germany}
 
\author{B. Argence}
 \affiliation{Laboratoire Kastler Brossel, Sorbonne Université, ENS-Université PSL, CNRS, Collège de France, 4 place Jussieu, 75252 Paris, France}
 
\author{G. Gredat}
 \affiliation{Université Paris-Saclay, CNRS, ENS Paris-Saclay, CentraleSupélec, LuMIn, Gif-sur-Yvette, France}
 
\author{F. Bretenaker}
 \affiliation{Université Paris-Saclay, CNRS, ENS Paris-Saclay, CentraleSupélec, LuMIn, Gif-sur-Yvette, France}
 
\author{N. Treps}
\email{nicolas.treps@lkb.upmc.fr}
 \affiliation{Laboratoire Kastler Brossel, Sorbonne Université, ENS-Université PSL, CNRS, Collège de France, 4 place Jussieu, 75252 Paris, France}

\date{\today}

\begin{abstract}
The noise dynamics of an Optical Frequency Comb (OFC) based on a mode-locked Ti-Sapphire laser is analyzed in terms of noise modes. A spectrally resolved multipixel homodyne detection enables the simultaneous measurement of the amplitude and phase noises of several optical frequency channels, from which the covariance matrices of the amplitude and phase quadratures of the laser field are calculated. The decomposition of these matrices into the four most significant time/frequency modes of the field enables the tracking of the origin of the noises and the correlations between the noise modes. In particular, the correlations between the amplitude and phase noises are measured. These measurements are well reproduced by a model taking into account the correlations between the CEO phase noise and the amplitude noise induced by the group velocity dispersion of the laser cavity.
\end{abstract}

\maketitle

\section{Introduction}
Mode-locked femtosecond lasers, or optical frequency combs (OFCs), have become an ubiquitous tool in metrology for the past 20 years. 
They were originally used to transfer the excellent spectral properties of the optical atomic clocks to the microwave frequency domain \cite{diddams2000direct,udem2002}.
They have now found applications in numerous fields, such as tests of fundamental physics \cite{Udem2001,Shelkovnikov2008,Fleurbaey2018,Grinin2020}, atomic and molecular spectroscopy \cite{Diddams2007,Picque2019}, time or frequency transfer \cite{ giorgetta_optical_2013,Guena2017, Predehl2012}, ranging measurements \cite{coddington2009rapid,Jian2012} or astrophysics \cite{ McCracken:17}. Many of those applications rely on the stability of the OFC. Furthermore, understanding the dynamics of OFCs, i.e. the noises affecting them, is critical to develop ultra-stable sources so that they do not limit the precision of the measurement.\\
For a single frequency laser, the dynamics is described in terms of amplitude noise (i.e. variation in the photon number), and phase noise (i.e. variation in frequency). However, for an OFC, composed of roughly 10$^5$ spectral lines, amplitude and phase noises affect each of them individually. Thus, such a complete characterisation is almost impossible due to the gigantic number of degrees of freedom. Yet, it has been theorised that the main dynamics of an OFC can be reduced to four distinct parameters: the pulse energy, the carrier envelope offset (CEO), the repetition rate and the central wavelength \cite{Haus1990, Haus1993}, indicating the presence of correlations between the different spectral lines. Each parameter is associated to a specific time/frequency mode of the electric field.\\
The dynamics of OFCs has been widely studied by measuring the noise on each parameter separately. A complete description of the noise in passively mode-locked lasers has first been developed by Haus and Mecozzi \cite{Haus1993} and later generalised by Paschotta \cite{paschotta2004noise1, paschotta2004noise2, paschotta2006optical}. A thorough  understanding of the dynamics of mode-locked fiber lasers is given in \cite{newbury2005theory, newbury_low-noise_2007}, both theoretically and experimentally. The intensity related dynamics of OFCs has drawn a lot of attention. Changes in the lasers parameters under pump laser modulation is studied in \cite{holman2003detailed, menyuk2007pulse, wahlstrand2007quantitative}. Those studies indicate that the main source of noise is the intensity fluctuations of the pump laser. They are responsible for several features such as frequency pulling, timing jitter and phase noise.\\
In previous studies, different setups were used to access the noise of the different laser parameters. Characterising the intensity noise is the easiest task as a single photodiode is sufficient. The CEO noise is usually characterised by beating the comb spectrum with its frequency-doubled counterpart. The spectrum must be expanded beforehand in order to cover an octave \cite{jones2000carrier}. The timing jitter can be measured by various techniques such as heterodyne beat with an ultra-stable laser \cite{hou2015timing} or by self-heterodyne with a fiber delay line interferometer \cite{jung2015all, tian2020optical}. Furthermore, most of the studies concentrate on the lower part of the noise spectrum, from the Hz to tenth of kHz. In this frequency range, technical noises are dominant. On the other hand, spectral correlations started to draw attention to track the Raman response of materials \cite{glerean2020time} for example, but also to measure amplitude and phase noise correlations across a the spectrum of a frequency comb \cite{brajato2020bayesian}. In this work, building on the experimental scheme introduced in \cite{Schmeissner:14} based on the investigation of spectral correlations, we present an experiment to measure with a single setup the four noise parameters at the quantum limit, enabling the full characterisation of the laser dynamics. To do so a spectrally resolved multipixel homodyne detection is used to measure the covariance matrices of the amplitude and phase quadratures of the field. The fluctuations of the laser parameters are recovered by extraction of the noise in the corresponding time/frequency mode from those covariance matrices. We focus our analysis on the frequency range of the noise spectrum from 200 kHz to 4 MHz. This range has drawn less attention in the literature. However at those frequencies, the laser is expected to reach the quantum limit and can thus be used for precise measurements. In this study, the laser is found to be shot noise limited around 3 MHz. In addition, the simultaneous measurement of the amplitude and phase quadratures gives access to correlations between amplitude and phase noises. The resulting amplitude-phase correlation matrices can be studied through singular value decomposition. Thus, we confirm that those correlations are a signature of the intensity related dynamics of the laser. We analyse this dynamics thanks to a simple model explaining the intensity dependence of the CEO frequency. By comparing the measured CEO frequency noise and the model, we show that in our laser, the CEO intensity dynamics is induced by the spectrum center frequency fluctuations via the group velocity dispersion of the laser cavity.

\section{Modal description of the noise}
\subsection{The perturbed pulse}
\label{sec:modal_description_A}
Our approach consists in investigating the noise of a single pulse from an optical frequency comb whose complex electric field is written as
\begin{equation}
E(t) = \mathcal{E}_0 a(t) \mathrm{e}^{-i \omega_c t},
\label{eq.:field_pulse}
\end{equation}
where $\mathcal{E}_0$ is the single photon field amplitude \cite{grynberg2010introduction}, $\omega_c$ is the carrier frequency, and $a(t)$ the slowly-varying Gaussian envelope of the pulse. This analysis can then easily be extended to the train of pulses emitted by an OFC according to
\begin{equation}
\label{eq.:E_comb(t)}
E_{comb}(t) = \sum_n E(t - n \tau_r) \mathrm{e}^{-i n \Delta \phi_{CEO} },
\end{equation} 
where $\tau_r$ is the repetition rate of the laser and $\Delta \phi_{CEO}$ is the carrier envelope offset (CEO) phase. \\
The laser pulses undergo intensity and phase noise due to various sources such as pump laser intensity fluctuations, vibrations or temperature fluctuations. In most cases, the resulting dynamics of the laser can be reduced to four parameters, $\overrightarrow{p} = (\delta \epsilon, \ \delta \omega_c, \ \delta \tau_{ceo}, \ \delta \tau_r)$ where $\delta \epsilon$ stands for amplitude fluctuations, $\delta \omega_c$ for carrier frequency fluctuations, $ \delta \phi_{CEO} = \omega_c \delta \tau_{ceo}$ for CEO phase fluctuations and $\delta \tau_r$ for timing jitter. Note that $ \delta \phi_{CEO}$ represents the fluctuations the carrier envelope offset phase and should formally be noted $\delta \Delta \phi_{CEO}$. This notation being too heavy we keep $\delta \phi_{CEO}$ to represent the CEO fluctuations.  \\
Thus, the starting point of our modal description of the dynamics is to study a single pulse (\ref{eq.:field_pulse}) undergoing a perturbation of those four parameters \cite{de2019modal}. The perturbed pulse is therefore written
\begin{equation}
E(t,\overrightarrow{p}) = \mathcal{E}_0 \left(1 + \delta \epsilon \right) a \left(t - \delta \tau_r \right) \mathrm{e}^{-i \left(\omega_c - \delta \omega_c \right) \left( t - \delta \tau_{ceo} \right)}.
\end{equation}
The same expression can be obtained in the spectral domain by taking the Fourier transform of the previous expression,
\begin{equation}
\label{eq.: E(Omega, p)}
E(\Omega, \overrightarrow{p}) = \mathcal{E}_0 \left(1 + \delta \epsilon \right) a \left( \Omega - \delta \omega_c \right) \mathrm{e}^{i \left( \omega_c \delta \tau_{ceo} + \Omega \delta \tau_r \right)},
\end{equation}
where $\Omega = \omega- \omega_c$. As those fluctuations are small the above expression can be expanded at first order to obtain the fluctuating electric field,
\begin{eqnarray}
\delta E(\Omega) &=& E(\Omega, \overrightarrow{p}) - E(\Omega) \nonumber \\ 
&\simeq& \mathcal{E}_0 \left[ \delta \epsilon a(\Omega) - \delta \omega_c \frac{\partial a (\Omega)}{\partial \Omega} \right. \nonumber \\
&&\quad \left. + i \omega_c \delta \tau_{ceo} a(\Omega) + i \Omega \delta \tau_r a(\Omega) \vphantom{\frac{\partial a (\Omega)}{\partial \Omega}} \right].
\label{eq.:fluctuations}
\end{eqnarray}
Note that the notation $\delta$ as in $\delta E(\Omega)$ implies the dependence in $\overrightarrow{p}$, i.e. $\delta E(\Omega) = \delta E(\Omega, \overrightarrow{p})$.\\
In the following, we write $a(\Omega) = \alpha(\Omega) \mathrm{e}^{i \phi (\Omega)} = \alpha_0 u(\Omega) \mathrm{e}^{i \phi (\Omega)}$. $\alpha_0$ is the field amplitude so that $\alpha_0 = \sqrt{N_0}$ with $N_0$ the mean photon number, $u(\Omega)$ is the mean field mode normalised as $\int \ddroit \Omega \vert u(\Omega) \vert ^2 = 1$ and $\phi(\Omega)$ the spectral phase. For simplicity we choose it to be constant over $\Omega$ and equal to zero, i.e. $\mathrm{e}^{i \phi (\Omega)}=1$. Without this simplification, the definition of the quadratures, introduced in the next section, as the real and imaginary parts of the field proportional to the amplitude and the phase, would be less straightforward.

\subsection{Field quadratures fluctuations}

The field fluctuations can also be written in terms of quadratures of the electric field $x(\Omega)$ and $p(\Omega)$ respectively proportional to the real and imaginary parts of the field defined by 
\begin{equation}
E(\Omega) = \frac{\mathcal{E}_0}{2} \left[ x(\Omega) + i p(\Omega) \right]. 
\end{equation}
Thus we have
\begin{equation}
2 \delta E(\Omega) = \mathcal{E}_0 \left[ \delta x(\Omega) + i \delta p(\Omega) \right],
\end{equation}
with $\delta x(\Omega) = 2 \delta \alpha (\Omega)$, proportional to the amplitude fluctuations, and $\delta p(\Omega) = 2 \alpha (\Omega) \delta \phi(\Omega)$, proportional to the phase fluctuations. By identification with eq. (\ref{eq.:fluctuations}) we find
\begin{eqnarray}
\delta x(\Omega) &=& 2 \alpha_0 \left[ \delta \epsilon u(\Omega) - \delta \omega_c \frac{\partial u(\Omega)}{\partial \Omega} \right],  \label{eq.:deltaX0}\\
\delta p(\Omega) &=& 2 \alpha_0 \left[ \omega_c \delta \tau_{ceo} u(\Omega) +  \Omega \delta \tau_r u(\Omega) \right].  \label{eq.:deltaP0}
\end{eqnarray}
In the expressions above, each of the four parameters is associated to a particular spectral mode related to the mean field one. Let us consider a Gaussian mean field spectral mode $u(\Omega) = \frac{1}{\sqrt[4]{2 \pi \Delta \omega^2}} \exp \left(- \frac{\Omega^2}{4\Delta \omega^2} \right)$, where $\Delta \omega$ is the spectral width of the field given by $\Delta \omega^2 = \int \ddroit \Omega^2 \vert u(\Omega)\vert^2 $. The quadratures (\ref{eq.:deltaX0}, \ref{eq.:deltaP0}) can finally be written as
\begin{eqnarray}
\delta x(\Omega) &=& 2 \alpha_0 \left[ \delta \epsilon u_{amp}(\Omega) - \frac{\delta \omega_c}{2 \Delta \omega} u_{cent-freq}(\Omega)\right], \label{eq.:deltaX}\\
\nonumber\\
\delta p(\Omega) &=& 2 \alpha_0 \left[ \omega_c \delta \tau_{ceo} u_{ceo}(\Omega) +  \Delta \omega \delta \tau_r u_{rep-rate}(\Omega) \right], \label{eq.:deltaP}\nonumber \\ 
\end{eqnarray}
with $u_{amp}(\Omega) \equiv u_{ceo}(\Omega) = u(\Omega)$, $u_{cent-freq}(\Omega) = -2 \Delta \omega \frac{\partial u(\Omega)}{\partial \Omega}$ and $u_{rep-rate}(\Omega) = \frac{\Omega u(\Omega)}{\Delta \omega}$. Those normalised spectral modes are named detection modes.\\
In conclusion, the dynamics of the laser can be recovered by accessing spectral modes in the amplitude and phase quadratures of the electric field. As a consequence, a spectrally resolved detection of the quadratures is needed. This is provided by a multipixel homodyne detection described in the next section. 

\section{Measuring the multimode field}

\begin{figure*}
\includegraphics[scale=0.6]{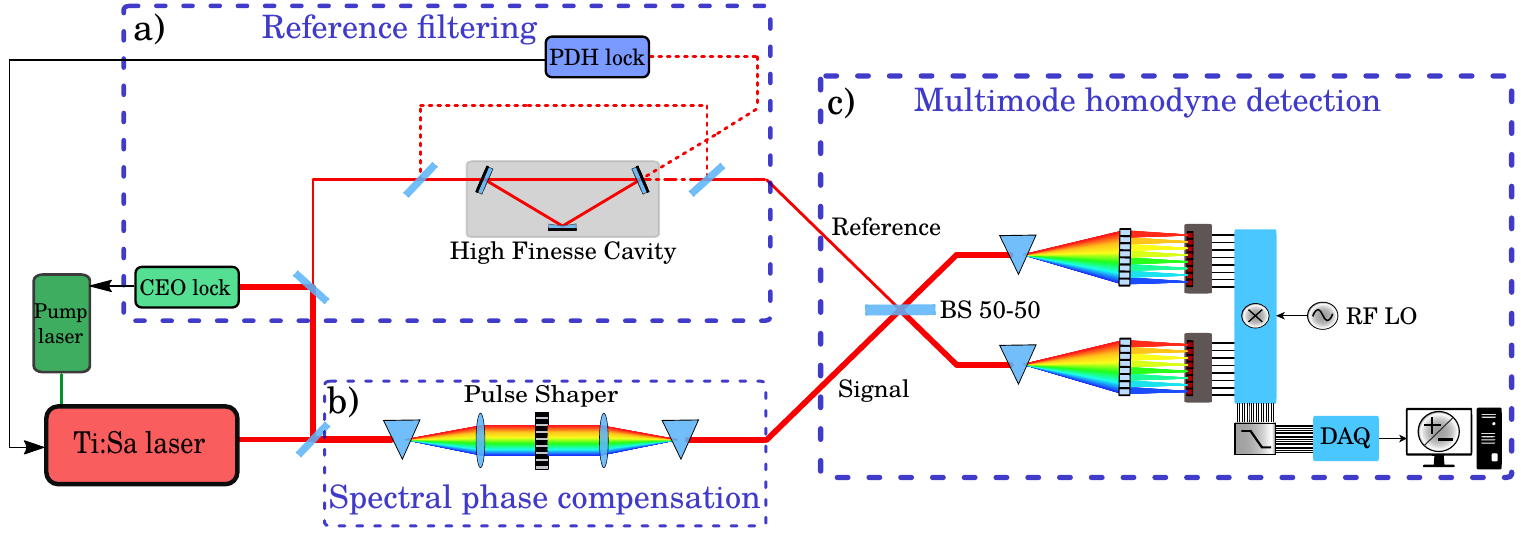}
\caption{\textbf{Experimental setup:} The experiment is composed of three parts. a) Reference filtering, to generate the reference beam via a high finesse cavity on which the repetition rate of the laser is locked. The CEO is also locked thanks to a f-2f interferometer to ensure maximum transmission through the cavity. The laser is free running above the locking bandwidth of a few kHz. b) Spectral phase compensation with a pulse shaper to ensure a flat phase for all spectral bands. That way, the homodyne is locked on the phase quadrature for all the spectral bands. c) Multipixel homodyne detection to measure the amplitude and phase noise in 8 different spectral bands. A demodulation stage is used to acquire noise at a given offset frequency selected by the radio-frequency local oscillator (RF LO) whose frequency can be swept. All the data acquired by an acquisition card (DAQ) and are post processed on a computer to recover the covariance matrices.}
\label{fig:manip}
\end{figure*}

The experiment is based on a spectrally resolved homodyne detection represented in Figure \ref{fig:manip}c. The laser spectrum is split in 8 spectral slices thanks to a grating and an array of microlenses.  Two detectors, composed of 8 photodiodes each, can then measure the amplitude and phase quadratures in each of those spectral bands. Consequently, the quadratures are measured in a pixelised basis with each pixel corresponding to a frequency band. We define the pixel basis $\left\{ v_i(\Omega) \right\}$, $i \in \left\{1:8\right\}$ with
\begin{equation}
    v_i(\Omega) =  \left\{
    \begin{array}{ll}
        u(\Omega_m) & \ \mathrm{if} \ \Omega_i \leq \Omega < \Omega_{i+1}, \\
        0 & \ \mathrm{otherwise,}
    \end{array}
\right.
\end{equation}
with $\Omega_m = \left( \Omega_i + \Omega_{i+1} \right)/2$. \\
The $x$ and $p$ quadratures of the field in this basis can be described by vectors $\overrightarrow{x} = \left(x_1, \ldots, x_8 \right)$, $\overrightarrow{p} = \left(p_1, \ldots, p_8 \right)$ with
\begin{eqnarray}
\label{eq.:pixelisation}
    x_i &=& \int v_i (\Omega) x (\Omega) \ddroit \Omega \ , \\
    p_i &=& \int v_i (\Omega) p (\Omega) \ddroit \Omega \ .
\end{eqnarray}
This scheme allows us to measure the spectral covariance matrices in amplitude $\Gamma^x$ with $ \Gamma^x_{ij} = \langle \delta x_i \delta x_j\rangle$ and in phase $\Gamma^p$ with $\Gamma^p_{ij} = \langle \delta p_i \delta p_j \rangle$ from which the dynamics of the laser can be extracted. The following section presents in details the experimental setup used and represented in Figure \ref{fig:manip}. 

\subsection{Experimental setup}

The laser under study is a Titanium-Sapphire based femtosecond oscillator from Femtolasers company. It delivers 22 fs FWHM pulses with a repetition rate of 156 MHz resulting in a 40 nm FWHM wide spectrum centered at 795 nm. The average power is of the order of 1 W. This laser is pumped by a 5W Finesse Pure CEP pump laser at 532 nm from Laser Quantum. The beam under study is first splitted in two: an intense beam, the signal and a weak one, the reference. The reference is filtered by a high finesse cavity (F$\simeq$1200) in order to decorrelate the high frequency phase noises of the two beams. This cavity acts as a low pass filter for the phase fluctuations with a bandwidth of approximately 125 kHz. To ensure maximum transmission through the cavity, the CEO frequency and the repetition rate of the laser are locked. To do so, the CEO frequency is first measured thanks to a f-2f interferometer and mixed with a radio-frequency reference signal to obtain an error signal. This signal is then used to stabilize the CEO frequency by acting on the pump laser current via a commercial PI servo controller from New Focus with a bandwidth of 3 kHz. The repetition frequency is stabilized by locking the laser cavity length on the high finesse cavity via a Pound-Drever-Hall (PDH) locking scheme. The PDH signal is derived from the reflected beam at the output of the cavity to avoid any modulation on the beam used for the detection. The error signal is fed to a mirror mounted on a piezoelectric actuator in the laser cavity via another PI servo controller with a bandwidth of 1 kHz. A detailed description of this locking scheme can be found in \cite{Schmeissner:14}. As both locks have a bandwidth of a few kHz, the laser is thus free running above those frequencies. Spectral phase compensation using a pulse shaper is used to ensure a flat phase across the 40 nm spectrum of the signal beam at the detection. This phase compensation is performed by a spatial light modulator LCOS-SLM X10468 (Liquid Crystal On Silicon) from Hamamatsu in a 4f configuration \cite{Monmayrant_2010}. The reference and signal beams are then recombined on a 50-50 beam splitter (BS). Finally, the spectrum is spatially spread and sent to two multipixel detectors thanks to arrays of microlenses. Each detector is composed of 8 photodiodes. The detected signals are split into a low frequency component (DC) and a high frequency one (AC), with a cutoff frequency around 200 kHz. The DC part is used for alignment purpose and to phase lock the homodyne detection on the phase quadrature. The 16 AC signals are mixed with a reference signal delivered by a frequency generator whose frequency is swept, and low pass filtered with a cutoff frequency of 10 kHz. The resulting signals are then acquired by a data acquisition card NI-PXIe 6361 from National Instrument with a sampling rate of 100 kSa/s. The demodulation stage is used to measure noises at higher frequencies than the 1 MHz bandwidth of the acquisition card but also to prevent the saturation of its dynamics. The laser being very noisy at low frequencies, acquiring all the noise spectrum in a single measurement would degrade the resolution of the spectrum at high frequencies where the noise is much lower. This scheme allows us to measure the covariance matrices at a given offset frequency set by the frequency of the demodulating signal. The data are then processed by a computer. This processing and normalisation procedure to retrieve the covariance matrices are detailed in the supplementary.

\subsection{Amplitude and phase covariance matrices}

\begin{figure*}
\includegraphics[scale=0.4]{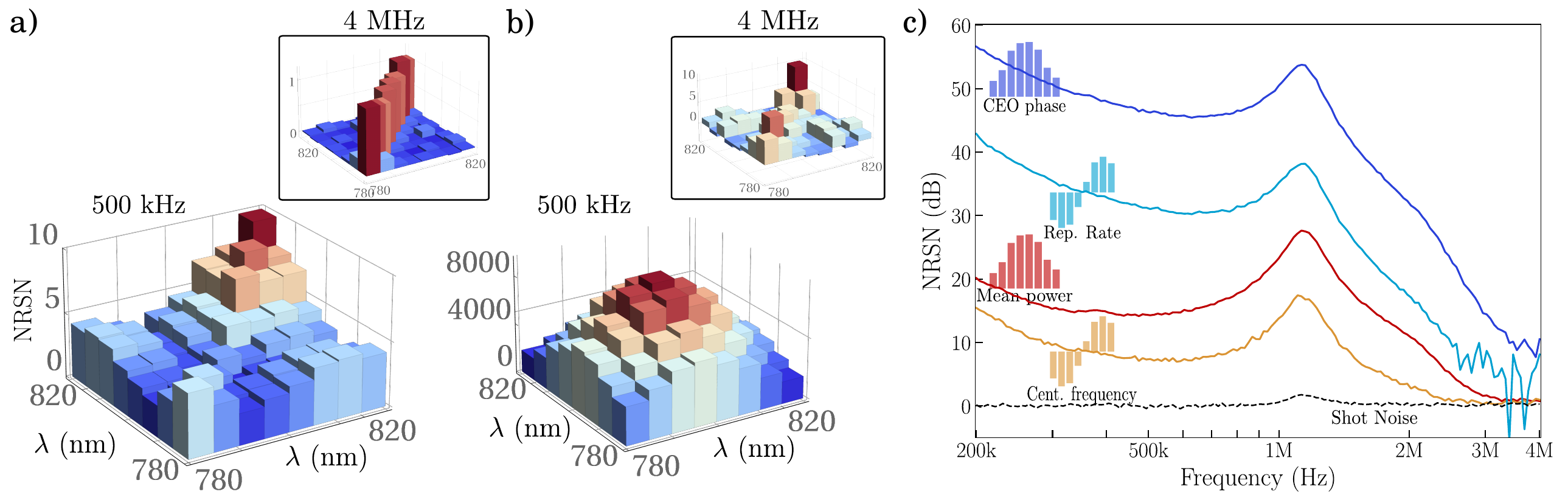}
\caption{\textbf{Experimental covariance matrices}. a) Amplitude and b): phase covariance matrices. The matrices are measured at 500 kHz. The noise is expressed as Noise Relative to the Shot Noise (NRSN) on a linear scale.  Correlations can be seen between the different spectral bands. In inset, matrices for noises at 4 MHz. At that frequency, the laser is only affected by the shot noise. c): noise on the physical parameters relative to the shot noise. Projection of the covariance matrices on the modes corresponding to the noise on the CEO, central and repetition rate frequencies as well as the mean power. The noises are expressed in dB relative to the shot noise. The shot noise is measured by blocking the reference beam and taking the difference of the photocurrents.}
\label{fig:cov_matrices}
\end{figure*}

An example of matrices acquired this way is shown in Figure \ref{fig:cov_matrices}. Figures \ref{fig:cov_matrices}$a$ and \ref{fig:cov_matrices}$b$ reproduce the covariance matrices for the amplitude and phase fluctuations, respectively, for an offset frequency of 500 kHz. The inset displays the same matrices for an offset of 4 MHz. The matrices are expressed in units of shot noise (noise relative to shot noise, NRSN), which is the standard quantum limit in sensitivity for amplitude and phase noise. The shot noise is a white and uncorrelated noise corresponding to the level of noise associated to a coherent state. Therefore, the level of noise displayed by the covariance matrices must be understood as an excess of noise compared to a coherent state of same mean power. \\
The amplitude covariance matrix measured at 4 MHz is diagonal with elements equals to one. It proves that the laser is only affected by the shot noise at this offset frequency. In this case, the laser filed can be approximated by a coherent state. On the contrary the amplitude and phase matrices at 500 kHz display correlations between spectral bands indicating that classical noise affects the laser dynamics. The next section presents how some information is extracted from those matrices.  \\
It can be noted that the phase matrix at 4 MHz is not purely diagonal as it is the case for the amplitude one. This is a result of the limited sensitivity of the detection for the phase noise. Indeed, in a standard homodyne detection, the phase measured is the one of the weak beam. However in our experiment, the weak beam is filtered and we aim at measuring the phase noise of the intense beam. Consequently, the normalisation introduces losses in the detection resulting in a decreased sensitivity as explained in the supplementary \ref{supp:normalisation}. 

\section{Extracting the noise}

\begin{figure*}
\includegraphics[scale=0.6]{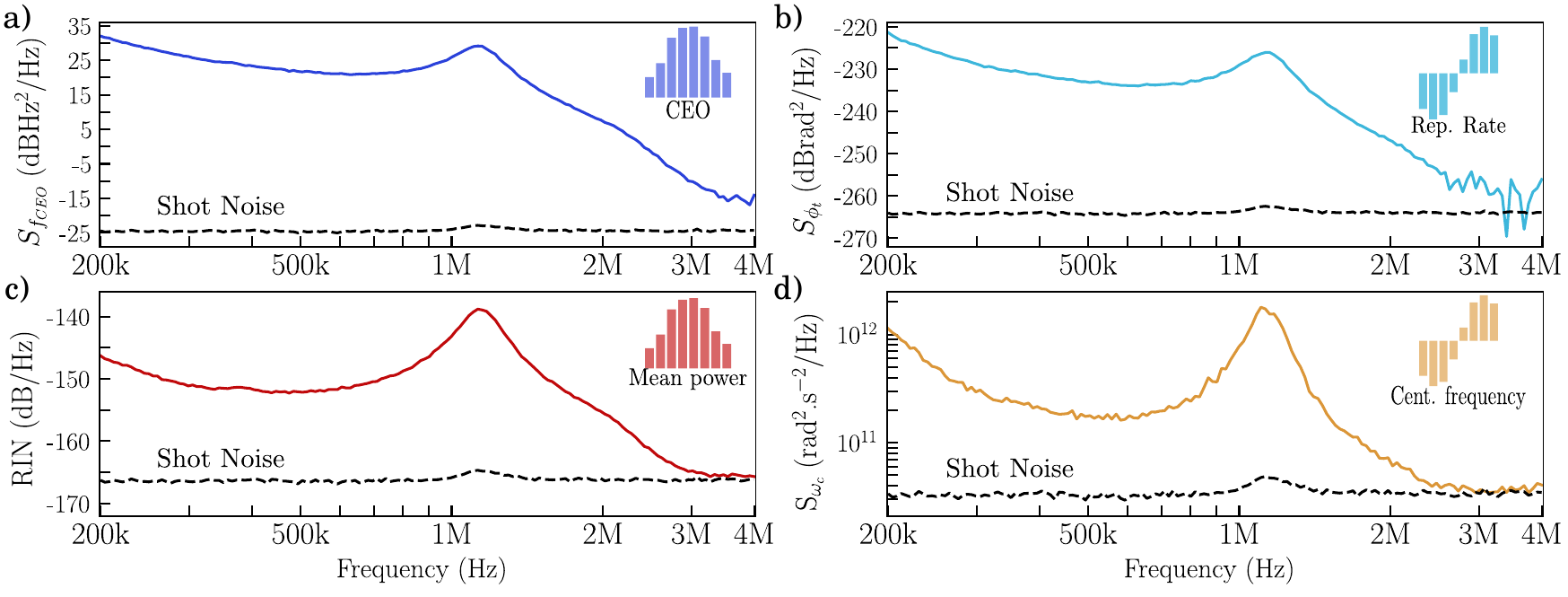}
\caption{\textbf{Noise in physical units} a) CEO frequency noise, b) timing jitter noise, c) mean power noise and d) spectrum center frequency noise. The spectral mode corresponding to each quantity is reproduced in each plot. The shot noise is plotted as a dashed black line.}
\label{fig:noise_physical_units}
\end{figure*}

To retrieve the noise spectrum of the laser parameters, the covariance matrices are measured at offset frequencies from 200 kHz to 4 MHz by sweeping the frequency of the demodulating signal. We note $\Gamma^{x,p} (f)$ the amplitude and phase covariance matrices calculated at the offset frequency $f$. Measuring the covariance matrix is enough to recover all the information on the dynamics assuming that the noises under study are Gaussian, which is a reasonable assumption in our experiment. Consequently, the noise relative to the shot noise for each of the four laser parameters, $(\delta \epsilon, \delta \omega_c, \delta \phi_{ceo} = \omega_c \delta \tau_{ceo}, \delta \tau_r$), defined in section \ref{sec:modal_description_A}., can be extracted by mathematically projecting the covariance matrices on the corresponding modes defined in equations (\ref{eq.:deltaX}, \ref{eq.:deltaP}). To do so, the noise modes need to be pixelised similarly to (\ref{eq.:pixelisation}) for quadratures. We note $\overrightarrow{u}_{mode}$ those modes. They are represented in Figure \ref{fig:cov_matrices}c. The resulting noise spectra are given by 

\begin{eqnarray}
\label{eq.:noise_relative_SN}
\nonumber
\delta \epsilon (f) &=& \left[ \overrightarrow{u}_{amp}^T \cdot  \Gamma^{x} (f) \cdot \overrightarrow{u}_{amp} \right]^{1/2},\\ \nonumber
\delta \omega_c (f) &=&  \left[\overrightarrow{u}_{cent-freq}^T \cdot  \Gamma^{x} (f) \cdot \overrightarrow{u}_{cent-freq}\right]^{1/2}, \\ \nonumber
\delta \phi_{ceo}(f) &=&  \left[\overrightarrow{u}_{ceo}^T \cdot \Gamma^{p} (f) \cdot \overrightarrow{u}_{ceo}\right]^{1/2}, \\ \nonumber
\delta \tau_r (f) &=&  \left[\overrightarrow{u}_{rep-rate}^T \cdot \Gamma^{p} (f) \cdot \overrightarrow{u}_{rep-rate}\right]^{1/2}, \\
\end{eqnarray}

where $\overrightarrow{u}^T_{mode}$ is the transposed mode.\\
As the covariance matrices are normalised to the shot noise, all the fluctuations in the expressions above are expressed in units of shot noise. Figure \ref{fig:cov_matrices}$c$ represents those fluctuations in dB relative to the shot noise as a function of the offset frequency, thus corresponding to the spectrum of the noise of each parameter. It can be seen that the dominant noises in this frequency range are the phase ones and mainly the noise on the CEO phase. The laser reaches the shot noise level around 3 MHz meaning that technical noises are no longer affecting the laser. The peak around 1 MHz corresponds to the relaxation oscillations of the laser. One can notice a discrepancy with respect to the shot noise at high frequency for the phase noises even though the phase noise is expected to reach the shot noise level at high frequencies. It is the consequence of the limited sensitivity of the measurement due to the renormalisation, as explained before.\\
To further investigate the dynamics of the laser, the noise spectra represented in Figure \ref{fig:cov_matrices}$c$ can be converted into physical units. From equations (\ref{eq.:deltaX}, \ref{eq.:deltaP}) and from the expressions of the NRSN (\ref{eq.:noise_relative_SN}), we thus obtain the following expressions for power spectral densities of the Relative Intensity Noise ($RIN(f)$),  the central frequency noise ($S_{\omega_c}(f)$), the CEO frequency noise ($S_{f_{CEO}}(f)$) and the timing phase noise ($S_{\phi_t}(f)$), with $\phi_t (f) = 2 \pi f_{r} \delta \tau_r (f)$, together with their units:
\begin{eqnarray}
\label{eq:physical_units}
\nonumber
RIN (f) &=&  \left( \frac{\delta \epsilon (f)}{\sqrt{N_0}} \right)^2 T_m  \quad \left[ \mathrm{Hz}^{-1} \right],\\ \nonumber
S_{ \omega_c} (f) &=& \left( \frac{\Delta \omega}{\sqrt{N_0}} \delta \omega_c (f) \right)^2 T_m \quad \left[ \mathrm{rad}^2 . \mathrm{s}^{-2}/\mathrm{Hz} \right],\\ \nonumber
S_{f_{CEO}} (f) &=& \left( \frac{f_r}{4 \pi \sqrt{N_0}} \delta \phi_{ceo}(f) \right)^2 T_m  \quad \left[ \mathrm{Hz}^2/ \mathrm{Hz} \right], \\ \nonumber
S_{ \phi_t} (f) &=& \left( \frac{\pi f_r }{\Delta \omega \sqrt{N_0}} \delta \tau_r (f) \right)^2 T_m \quad \left[ \mathrm{rad}^2/ \mathrm{Hz} \right],\\
\end{eqnarray}
where $N_0$ corresponds to the number of photons hitting the detector during the acquisition duration. It is given by $N_0 = P  T_{m} / \hbar \omega_c $ where $P$ is the optical power before the BS of the homodyne detection ($P = 11$ mW), and $T_{m} = 1/BW$ the acquisition time with $BW$ the bandwidth of the low-pass filter used in the detection chain after the demodulation ($BW = 10$ kHz). The spectra obtained are reproduced in Figure \ref{fig:noise_physical_units}. As a comparison, the timing phase noise and the RIN of a mode-locked laser have been determined theoretically in \cite{paschotta2004noise2} where similar noise levels have been found. Those spectra offer a quantitative measurement of the noise affecting the OFC, derived from a single measurement. These quantities are used in section \ref{sec:model_fceo} to track the origin of the noises, in particular the intensity related dynamics of the CEO frequency.

\section{Unveiling the dynamics from XP correlations}
\label{sec:schmidt}
\begin{figure*}
\includegraphics[scale=0.3]{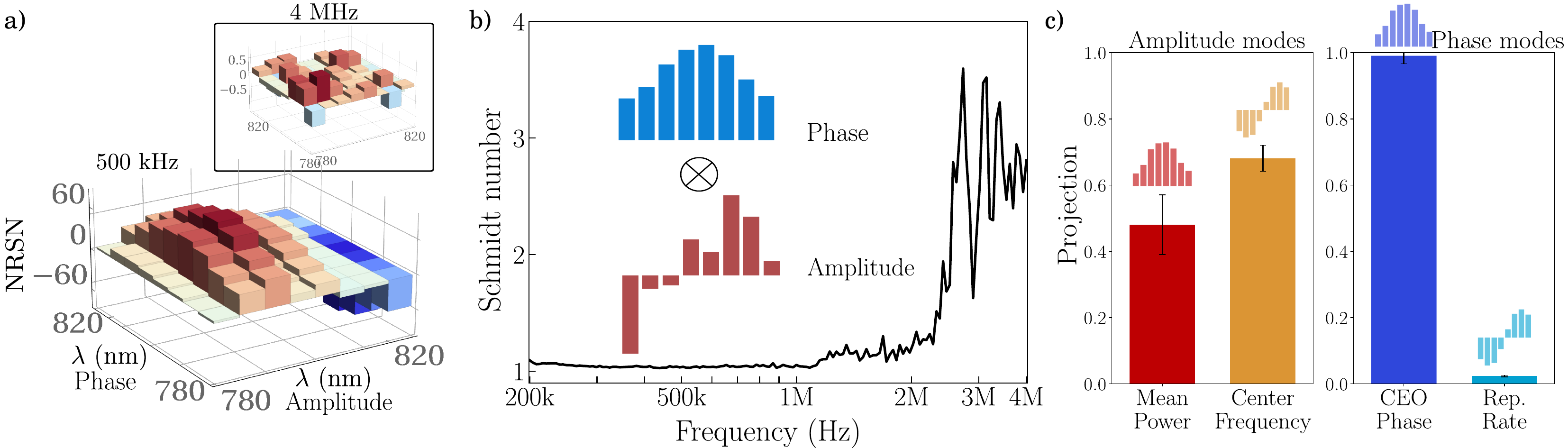}
\caption{\textbf{Phase and amplitude correlations:} a) Correlation matrix between amplitude and phase at an offset frequency of 500 kHz. In inset the same but at 4 MHz where the correlations vanish. b) Schmidt decomposition of the correlations matrices. The Schmidt number as a function of the offset frequency is plotted together with the amplitude and phase singular modes for the frequencies where it is equal to one. c) Projection of the amplitude and phase singular modes on the detection modes.}
\label{fig:corr_XP}
\end{figure*}

The experimental scheme allows the measurement of the amplitude and the phase quadratures simultaneously by taking respectively the sum and difference of the photocurrents of the 16 photodiodes after acquisition. Consequently, it is also possible to access the correlation matrix $ \Gamma_{ij}^{xp} = \langle \delta x(\Omega_i) \delta p(\Omega_j) \rangle$ between amplitude and phase. An example of such matrix is represented in Figure \ref{fig:corr_XP}$a$ for an offset frequency of 500 kHz (4 MHz in inset). 

These matrices show spectral correlations between the amplitude and the phase up to the MHz domain and no correlations above. To analyse this matrix, a singular value decomposition (SVD) is performed. From this decomposition the Schmidt number can be calculated. It is given by $K = \left( \sum_n \lambda_n^2 \right)^2 / \sum_n \lambda_n^4$, where the $\lambda_n$'s are the singular values \cite{christ2011probing}. This parameter provides some information on the number of significant modes involved in the process. It is represented in Figure \ref{fig:corr_XP}$b$ as a function of the offset frequency. The Schmidt number is equal to one up to 1 MHz meaning that only one pair of modes, one in phase and one in amplitude, is necessary to reconstruct the correlations. Those two modes, represented in the same Figure, are the singular modes. Because only one pair of mode is involved we can assume that only one noise source is responsible for those correlations. As it has been shown in various papers \cite{holman2003detailed, wahlstrand2007quantitative} it is expected to come from the fluctuations of the pump laser intensity. This is investigated in the next part.\\
A qualitative understanding of the underlying processes can be obtained through the projection of the singular modes on the detection modes introduced earlier. The projection of the amplitude singular mode on the mean power and spectrum center frequency detection modes and the projection of the phase singular mode on the CEO and repetition rate detection modes are represented in Figure \ref{fig:corr_XP}$c$. It can be seen that mainly three modes are coupled. In phase, only the CEO detection mode is coupled to the amplitude ones. Indeed, this was expected as it is well-known that intensity fluctuations, induced by pump power fluctuations, have a huge impact on the CEO of the laser and are even used to control it \cite{xu_route_1996} as it is the case in this study. In amplitude, the singular mode is a linear combination of the mean power and spectrum center frequency modes. The dominant contribution comes from the fluctuations of spectrum center frequency. Despite the fact that there is a strong coupling between the CEO and the mean power fluctuations due to the Kerr effect, this contribution can be exceeded by the fluctuations of the spectrum center frequency. This is the case when there is a non negligible residual group velocity dispersion inside the laser cavity \cite{holman2003detailed}. In the next section we investigate those correlations with a simple model describing the intensity related dynamics of the CEO fluctuations. 

\section{Intensity related dynamics: Recovering the CEO fluctuations}

\label{sec:model_fceo}
As previously demonstrated, the dominant noise is the CEO frequency fluctuations. Moreover only one noise source is expected to induce correlations between amplitude and phase as suggested by the SVD applied in the previous section. As a consequence, following the idea developed in \cite{holman2003detailed}, the intensity related dynamics of the CEO frequency is investigated.\\
The CEO frequency is defined as follows
\begin{equation}
f_{CEO} = \frac{f_r}{2\pi} \Delta \phi_{CEO} = \frac{\omega_c}{2\pi} \left( 1 - \frac{v_g}{v_{\phi}} \right),
\end{equation}
where $\omega_c$ is the carrier frequency and $v_g$ and $v_{\phi}$ the average group and phase velocities defined by
\begin{align}
\frac{c}{v_g}& = \overline{n} + \omega_c \frac{\mathrm{d} \overline{n}}{\mathrm{d}\omega}\ ,\\
\frac{c}{v_{\phi}}& = \overline{n}\ .
\end{align}
These velocities are averaged over the cavity length and contain the refractive index  $\overline{n} = \overline{n}_0 + \overline{n}_2 I$, which is itself averaged over the cavity and includes the Kerr effect. An important point to note is that $\overline{n}$ depends on the intensity $I$ and on $\omega_c$, due to the dispersion,  which itself depends on $I$. The central frequency can be affected by intensity fluctuations when an asymmetry between the gain and the loss profiles exists \cite{Haus1993, menyuk2007pulse}. In that configuration a change in the gain results in a frequency pulling effect due to a shift of the equilibrium frequency.  \\
Thus, the intensity dependence of the CEO frequency is given by
\begin{eqnarray}
\frac{\ddroit f_{CEO}}{\ddroit I} &=& \frac{1}{2 \pi} \left[ \frac{\ddroit \omega_c}{\ddroit I} \left( 1 - \frac{v_g}{v_{\phi}} \right) \right. \nonumber \\ 
 &+& \left. \ \omega_c \frac{v_g}{v_{\phi}}\left( \frac{1}{v_{\phi}} \frac{\ddroit v_{\phi}}{\ddroit I} - \frac{1}{v_g} \frac{\ddroit v_g}{\ddroit I} \right) \right]. \label{eq.:dfceo/dI}
\end{eqnarray}
The model is derived with respect to the intra-cavity peak intensity $I$, which we calculate for a Fourier transformed limited pulse (the effect of dispersion is included later). However, in practice, we measure the fluctuations of the parameters with respect to the laser mean output power $P$. The conversion from $I$ to $P$ is given by
\begin{equation}
\frac{\ddroit I}{\ddroit P} = \frac{2 \times 0.88}{f_r T_{coupler} \Delta t_{pulse} \pi w^2}.
\label{eq.:dIdP}
\end{equation}
This quantity is determined from the parameters of the experiment: $f_r = 155$ MHz, $T_{coupler} = 0.28$ the transmission of the output coupler, $\Delta t_{pulse} = 22$ fs the pulse duration and $w = 10$ $\mu$m the waist in the crystal. The factor 0.88 comes from the hyperbolic secant shape of the pulse. Given those data we find $\ddroit I / \ddroit P \simeq 5.8 \cdot 10^{-15}$ m$^{-2}$.\\
From equations (\ref{eq.:dfceo/dI}) and (\ref{eq.:dIdP}), the CEO frequency fluctuations are given by

\begin{figure}
\includegraphics[scale=0.44]{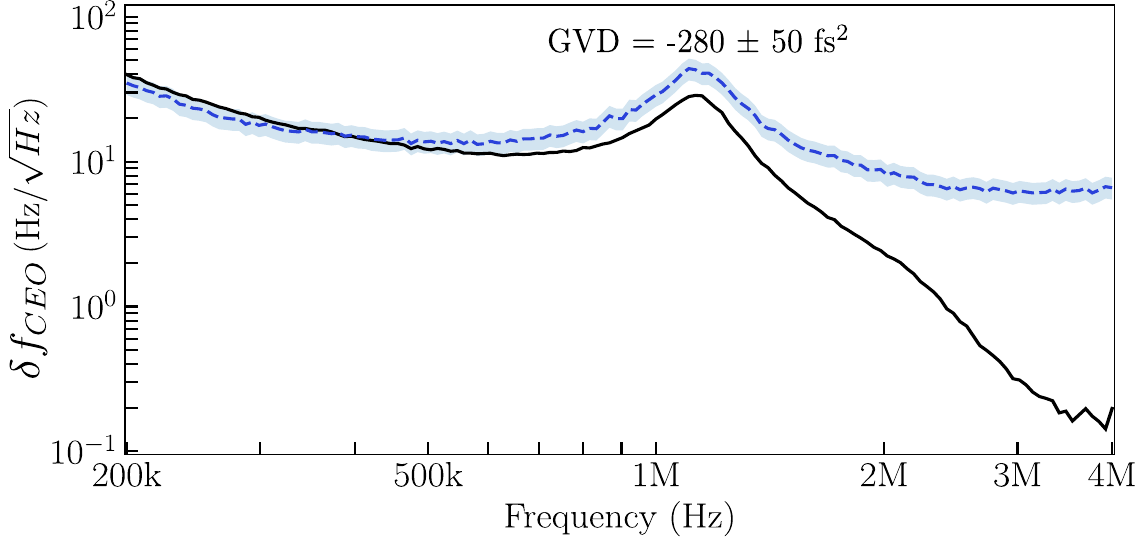}
\caption{\textbf{Experimental versus model for the CEO noise:} In dashed blue the model from (\ref{eq.:f_ceo_model}), the shaded zone corresponds to the uncertainty on the GVD determined in the laser. Plain line: measured CEO frequency fluctuation given by $\delta f_{CEO} = \sqrt{S_{f_{CEO}}}$ from (\ref{eq:physical_units}).}
\label{fig:f_ceo_model}
\end{figure}

\begin{widetext}
\begin{equation}
    \delta f_{CEO} = \frac{1}{2\pi} \left[ \delta \omega_c \left( \left( 1-\frac{v_g}{v_{\phi}} \right) + \omega_c \frac{v_g^2}{v_{\phi}} \mathrm{GVD} - \omega_c \frac{v_g}{c} \frac{\partial \overline{n}}{\partial \omega} \right)  + \delta P \frac{\ddroit I}{\ddroit P} \frac{\omega_c v_g}{c} \left( \frac{v_g}{v_{\phi}} \omega_c \frac{\partial \overline{n}_2}{\partial \omega} - \overline{n}_2 \left( 1-\frac{v_g}{v_{\phi}} \right) \right) \right],
\label{eq.:f_ceo_model}
\end{equation}
\end{widetext}
where GVD is the average Group Velocity Dispersion inside the cavity given by $\mathrm{GVD} = \frac{\ddroit}{\ddroit \omega} \left( \frac{1}{v_g} \right)$.\\
From this expression it is clear that the noise on the CEO frequency arises from the noise on the central frequency $\delta \omega_c$ and from the mean power fluctuations $\delta P$. The factor coupling the CEO frequency and the central frequency is composed of three terms. The first one comes from the dispersion in the laser, $\left( 1 - \frac{v_g}{v_{\phi}} \right)$, the second one from the group velocity dispersion, GVD, and the last one is due to the dispersion of the Kerr effect $\frac{\partial \overline{n}}{\partial \omega}$. On the other hand, the term coupling the CEO frequency to the intensity is mainly due to the Kerr effect, as expected, via $\overline{n}_2$ and its dispersion. \\
To apply this formula, a few quantities need to be taken from the literature. We have $\overline{n}$ = 1.00116, $\overline{n}_2$ = 1.8$\cdot 10^{-23} \ $m$^2$W$^{-1}$, $\frac{\partial \overline{n}_0}{\partial \omega}$ = 3.5$\cdot 10^{-21}$ s and $\frac{\partial \overline{n}_2}{\partial \omega}$ = 3$\cdot 10^{-39}$ sm$^2$W$^{-1}$ \cite{holman2003detailed}. Finally, the group velocity dispersion (GVD) of the laser needs to be estimated. Just after the output coupler, the duration of the pulse is 36 fs for a 40 nm spectrum, showing that the pulse is chirped. To have an estimation of the GVD inside the cavity, the passage through the output coupler needs to be taken into account. We estimate the dispersion introduced by 4 mm of silica of dispersion $\beta_2 = 36$ fs$^2$/mm. After the output coupler the pulse duration is found to be 24 fs. The passage through the crystal also needs to be taken into account. However, as we derived the expression (\ref{eq.:f_ceo_model}) of the fluctuations averaged over the cavity, this quantity varies depending on where the noise arises in the laser cavity. Consequently, an uncertainty can be associated to the estimation of the GVD. Because the cavity is linear, each pulse goes through the crystal twice per round-trip. Thus, to estimate the dispersion and its uncertainty we calculate the GVD for one trip in the crystal and calculate the uncertainty associated to a pulse which has not yet traveled through the crystal or which has done a double pass. The crystal is made of 3 mm of sapphire of dispersion $\beta_2 = 58$ fs$^2$/mm. The resulting calculated GVD is -280 fs$^2$ and the uncertainty is $\pm$ 50 fs$^2$. The calculation to estimate the GVD is detailed in the supplementary.\\ 
The model can now be applied to our measurements. We use the fluctuations of the mean power, $\delta P$, and of the central frequency, $\delta \omega_c$, measured experimentally thanks to our setup. For each offset frequency, the expected CEO frequency fluctuation is calculated using those values as well as the estimated GVD. This gives a spectrum for the CEO frequency fluctuations, which can be compared to the one measured with the setup. The resulting trace is reproduced in Figure \ref{fig:f_ceo_model} alongside of the measured CEO frequency fluctuations given by $\delta f_{CEO} = \sqrt{S_{f_{CEO}}}$. A good agreement is found between the model and the measured traces. The experimental data almost entirely fall in the uncertainty area up to 1 MHz. This agreement proves that the CEO dynamics is indeed related to the intensity fluctuations of the laser and that this feature is sufficient to explain it almost entirely. It confirms what was found with the SVD: the Schmidt number is equal to one as one noise source is responsible for the main dynamics.  It also demonstrates the coupling between the CEO and the central frequency fluctuations which was found by the projection of the singular modes. Nonetheless, the model does not seem accurate above 1 MHz. One explanation is that the noise on the central frequency is really low and close to the shot noise around 2 MHz, thus around that frequency the model cannot properly reproduce the CEO noise. This could also be due to some additional filtering coming from the laser cavity or the detection scheme not taken into account in the model. \\
To further explore the model, the ratio of the contributions of the center frequency and mean power fluctuations, calculated from the model, to the measured CEO fluctuations are reported in Table \ref{tab:contributions}. It can be seen that the dominant contribution is the one coming from the spectrum center frequency. More precisely it is the spectrum center frequency via the GVD which seems to be the dominant contribution. This is in agreement with the results found in Section \ref{sec:schmidt}. The center spectrum detection mode has a higher contribution to the amplitude singular mode as it is directly coupled to the CEO frequency fluctuations via the GVD. This decomposition indicates that the mean power fluctuations has a really small direct effect on the CEO frequency. Note that in Figure \ref{fig:corr_XP}$c$ there is still a significant contribution to the amplitude singular mode from the mean power because the fluctuations of center frequency are also coupled to the intensity fluctuations due to an asymmetry between the gain and loss profiles as explained before.  Thus, the intensity has a significant impact on the CEO frequency only when there is a residual group velocity dispersion inside the laser cavity. This has also been identified in \cite{holman2003detailed, newbury2005theory}. The knowledge of this processes can help to improve the performance of frequency combs. To achieve a lower CEO noise, the noise of the pump laser can be reduced or the GVD of the laser cavity can be reduced so that this intensity noise has a lower effect. In practice the first option is probably the easiest to implement. Alternately, a better lock of the CEO frequency can be achieved by using a laser cavity with an appreciable amount of group velocity dispersion.

\begin{table*}
\caption{\label{tab:contributions}
\textbf{Expressions and value of the ratio of contributions of the model to the measured CEO fluctuations}. This ratio is averaged over the frequency range where the model is the most accurate, from 200 kHz to 500 kHz. The first three lines are the contributions from the spectrum center frequency and the last one the contribution from the power fluctuations.
}
\begin{ruledtabular}
\begin{tabular}{llc}
\textrm{Contribution}&
\textrm{Expression}&
\textrm{ratio $ \vert c_j / \delta f_{CEO,exp} \vert $}\\
\colrule
\\
\vspace{2mm}
Group Velocity Dispersion & $c_1 =\frac{1}{2 \pi} \delta \omega_c \left( \omega_c \frac{v_g^2}{v_{\phi}} \mathrm{GVD} \right)$ & $9 \cdot 10^{-1}$\\
\vspace{2mm}
Kerr Effect Dispersion & $c_2 =\frac{1}{2 \pi} \delta \omega_c \left( \omega_c \frac{v_g}{c} \frac{\partial \overline{n}}{\partial \omega} \right)$ & $4 \cdot 10^{-2}$\\
\vspace{2mm}
Dispersion & $c_3 = \frac{1}{2 \pi} \delta \omega_c \left( 1 - \frac{v_g}{v_\phi} \right)$ & $5 \cdot 10^{-3}$\\
\vspace{2mm}
Kerr Effect & $c_4 =\frac{1}{2 \pi} \delta P \frac{\ddroit I}{\ddroit P} \frac{\omega_c v_g}{c} \left( \frac{v_g}{v_{\phi}} \omega_c \frac{\partial \overline{n}_2}{\partial \omega} - \overline{n}_2 \left( 1-\frac{v_g}{v_{\phi}} \right) \right)$ & $2 \cdot 10^{-6}$\\

\end{tabular}
\end{ruledtabular}
\end{table*}

\section{Conclusion}
We have presented a method enabling the characterisation of the dynamics of an optical frequency comb with a single setup. From the amplitude and phase covariance matrices, the fluctuations of the laser parameters have been measured at the shot noise limit. The noises are also expressed in physical units to be compared to other studies. The noise levels measured are similar to what can be expected from the literature on OFC dynamics. In addition, we measured the amplitude-phase correlation matrix. Thanks to a singular value decomposition, the underlying processes can be studied. We show that those correlations are mainly induced by the pump laser intensity fluctuations. Our analysis permits the identification of the coupling mechanisms. The fluctuations of the spectrum center frequency induced by the pump noise via frequency pulling is the main driving force of the CEO frequency noise. This is due to the group velocity dispersion of the laser cavity which was estimated to be $-280$ fs$^2$. \\
Using such setup could help for the design of low noise frequency combs by characterising their dynamics. This scheme could also be improved to implement a feedback on the laser pump current as it is the main source of noise. Another improvement is to use a fiber-optic delay line to decorrelate the phase noise of the two arms of the homodyne detection. This scheme is easier to implement and can give access to noise at lower frequencies where the technical noise can be investigated.

\begin{acknowledgments}
This work was supported by the Agence Nationale de la Recherche (LASAGNE ANR-16-ASTR-0010-03) and the Direction Générale de l'Armement.
\end{acknowledgments}

\bibliography{biblio}

\appendix
\section{Determination of the laser's dispersion}

The aim is to determine the GVD parameter in the laser. The starting point is equation (2.118) in \cite{weiner2011}.
\begin{equation}
\lbrace \left(D_g +iD \right) \frac{\ddroit ^2}{\ddroit t^2}+(g-l-i\psi)+(\gamma-i\delta) \vert a(t) \vert ^2 \rbrace a(t) = 0,
\label{eq.:master_eq}
\end{equation}
with $g$ the gain per roundtrip and $\Omega_g^2$ the width of the gain in amplitude so that $D_g = g/\Omega_g^2$, $l$ the losses per roundtrip, $\gamma$ the automodulation coefficient in amplitude, $\delta$ the self phase modulation coefficient, $D$ the group velocity dispersion (GVD) and $\psi$ the dephasing per roundtrip. We also use $U = 2 a_0^2 t_p$ the pulse energie as well as the normalised quantities $D_n = D/D_g$, $t_{pn} = Ut_p/2D_g$. \\
The solution of the equation for the mode-locked laser is a chirped hyperbolic secant
\begin{equation}
a(t) = a_0 \left[ \mathrm{sech}\left( \frac{t}{t_p} \right) \right] ^{1+i\beta}.
\label{eq.:sech}
\end{equation}
By putting (\ref{eq.:sech}) in (\ref{eq.:master_eq}) we find
\begin{equation}
\left\lbrace
\begin{array}{ccc}
&g-l-i \psi + \frac{ \left( 1+i \beta \right)^2}{t^2_p} \left( D_g + i D \right) = 0,\\
&\left( D_g + i D \right) \frac{2 + 3 i \beta - \beta^2}{t^2_p} = \left( \gamma - i \delta \right)a^2_0.
\end{array}\right.
\end{equation}
By multiplying the second equation by $D_g/t^2_p$ and identifying the real and imaginary part, we have:
\begin{equation}
\left\lbrace
\begin{array}{ccc}
& 2 - \beta^2 - 3 \beta D_n = \gamma t_{pn}\\
& 3 \beta + 2 D_n - \beta^2 D_n = - \delta t_{pn}
\end{array}\right.
\end{equation}
From the second equation we have the GVD depending on the duration of the pulse and its chirp.
\begin{equation}
D = D_n \times D_g = D_g\frac{-\left( \delta t_{pn} + 3 \beta \right)}{2 - \beta^2}
\end{equation}
Consequently, to determine the GVD, the duration of the pulse and its chirp in the cavity need to be known. At the output of the laser the pulse duration has a duration of 36 fs and a spectrum of 40 nm corresponding to a Fourier transform spectrum of 22 nm. Consequently, the chirp parameter of the pulse is $\beta_0 = \sqrt{(36/22)^2-1} = 1.3$. The chirp parameter is defined so that the electric field of the chirped pulse is given by: $E(t) = E_0 \mathrm{exp} \left( - \frac{1+i\beta}{2} \frac{t^2}{\Delta t_0}\right)$.\\
The evolution of the duration and chirp of a chirped pulse going through $z$ mm of a dispersing medium of dispersion $\beta_2$ is given by:
\begin{align}
\Delta t(z) &= \Delta t_0 \sqrt{\left(1+\frac{\beta_2 z \beta_0}{\Delta t_0} \right)^2 + \left( \frac{\beta_2 z}{\Delta t_0^2} \right)^2}\\
\beta(z) &= \beta_0 + \frac{\beta_2 z \beta_0^2}{\Delta t_0^2} + \frac{\beta_2 z }{\Delta t_0^2}
\end{align}

\section{Homodyne detection:}

The fields $E^{(+)}_{\pm}$ at each output of the beam-splitter (BS) can be expressed as
\begin{equation}
\label{eq.:BS_output}
E^{(+)}_{\pm}(t) = \frac{E^{(+)}_{s}(t) \pm E^{(+)}_{ref}(t)}{\sqrt{2}}
\end{equation}
Where $E^{(+)}_{s}(t) = \mathcal{E}_0 a_{s}(t) e^{-i \omega_0 t}$ is the field coming from the intense beam and $E^{(+)}_{ref}(t) = \mathcal{E}_0 a_{ref}(t) e^{-i \omega_0 t}$ from the weak one. The instantaneous intensity detected at each BS output, $S_{\pm}$, given according to $I(t) = 2 \epsilon_0 n c \vert E^{(+)}(t)\vert ^2$ in (J/s/m$^2$), is
\begin{eqnarray}
\label{eq.:BS_intensity}
S_{\pm}(t) &=& \frac{I_{s}(t)}{2} + \frac{I_{ref}(t)}{2} \nonumber \\
&\pm& nc\epsilon_0 \lbrace E^{(+)}_{s}(t)E^{(-)}_{ref}(t) + E^{(-)}_{s}(t)E^{(+)}_{ref}(t) \rbrace \nonumber \\
\end{eqnarray}
At this point it is more convenient to work in the frequency domain. The intensity at each output can thus be written
\begin{equation}
S_{\pm}= \frac{I_{s}}{2} + \frac{I_{ref}}{2} \pm \mathcal{E}_{0}^2 nc\epsilon_0 \lbrace a_{s}(\Omega)a_{ref}^{\star}(\Omega) + a_{s}^{\star}(\Omega)a_{ref} (\Omega) \rbrace
\end{equation}
The envelope of the field can be decomposed in a modulus and phase term
\begin{align}
\label{eq.:signal_decompo}
&a_{ref}(\Omega) = \vert a_{ref}(\Omega) \vert e^{i\phi_{ref}(\Omega)} = \alpha_{ref}(\Omega) e^{i\phi_{ref}(\Omega)}\\
\label{eq.:LO_decompo}
&a_{s}(\Omega) = \vert a_{s}(\Omega) \vert e^{i\phi_{s}(\Omega)} = \alpha_{s}(\Omega) e^{i\phi_{s}(\Omega)}
\end{align}

Where $\phi_j(\Omega)$ is the phase of the reference or signal pulse.

Summing the signal from both detectors gives access to an amplitude measurement of the signal. Indeed, because the signal field is stronger than the reference one, the intensity fluctuations measured are mainly the signal beam ones. The signal $\mathcal{I}_+ = S_+ + S_-$, can be written as
\begin{align*}
\mathcal{I}_+(\Omega) & = \mathrm{I}_s + \mathrm{I}_{ref}\\
& = 2 \epsilon_0 n c \mathcal{E}_0^2 \left( \vert a_{s}(\Omega) \vert ^2 + \vert a_{ref}(\Omega) \vert ^2 \right) \\
& = 2 \epsilon_0 n c \mathcal{E}_0^2 \left( \alpha_s^2(\Omega) + \alpha_{ref}^2(\Omega) \right)
\end{align*}

Consequently, the fluctuations of the sum of the signals is given by
\begin{equation}
\label{eq.:delt_I_+}
\delta \mathcal{I}_+(\Omega) \propto \alpha_{s}(\Omega)\delta \alpha_{s}(\Omega) \quad \Rightarrow \quad \delta \mathcal{I}_+(\Omega) \propto  \alpha_s(\Omega)\delta x_{s}(\Omega)
\end{equation}

Taking the difference of the signals leads to a measurement of the relative phase between the two fields.
With the decompositions (\ref{eq.:signal_decompo}) and (\ref{eq.:LO_decompo}), the difference of the output intensities is

\begin{equation}
\mathcal{I}_{-}(\Omega) = \alpha_s(\Omega)\mathcal{E}_{0}^2 nc\epsilon_0 \lbrace  \alpha_{ref}(\Omega) e^{i\phi(\Omega)}+  \alpha_{ref}(\Omega) e^{-i\phi(\Omega)} \rbrace
\end{equation}
Where $\phi(\Omega) = \phi_{s}(\Omega) - \phi_{ref}(\Omega)$ is the relative phase between the two pulses. \\
In what follows we are interested in studying the phase fluctuations of the signal. To do so the field's ($E^{(+)}(\Omega) = \mathcal{E}_0  \alpha(\Omega) e^{i \phi (\Omega)}$) fluctuations, $\delta E$, can be expanded to the first order:

\begin{equation}
\label{eq.:field_fluctuation}
\delta E^{(+)}(\Omega) = \mathcal{E}_0 e^{i \phi (\Omega)} ( \delta \alpha(\Omega) + i \delta \phi (\Omega) \alpha(\Omega) ) 
\end{equation}

As the signal is intense, the amplitude fluctuation can be neglected and the field expressed by its mean value $\alpha_{s}$. With expressions (\ref{eq.:field_fluctuation}), the fluctuations of the difference of output intensities can be written 
\begin{eqnarray}
\label{eq.:delt_I_-_all}
\delta \mathcal{I}_{-}(\Omega) =  2 \alpha_{s} \mathcal{E}_{0}^2 nc\epsilon_0 \lbrace \delta \alpha_{ref}(\Omega) \cos \phi(\Omega) \nonumber \\
 - \alpha_{ref}(\Omega) \delta \phi (\Omega) \sin \phi(\Omega) \rbrace
\end{eqnarray}
The phase difference between the reference and the signal needs to be the same for all frequencies and set to $\pm \frac{\pi}{2}$ by locking the relative path of the two arms. This locking point ensures to retrieve the phase fluctuations when the difference of the signals is taken. Finally, as we are interested in the signal phase fluctuations only, the phase fluctuations of the reference have to be low. This can be accomplished by filtering the reference field so that $\delta \phi (\Omega) = \delta \phi_{s}$.\\
Taking in account all those effects, the final signal is given by
\begin{equation}
\label{eq.:delt_I_-}
\delta \mathcal{I}_{-}(\Omega) \propto \alpha_{s}(\Omega) \alpha_{ref}(\Omega) \delta \phi_{s} (\Omega) \ \Rightarrow \ \delta \mathcal{I}_{-}(\Omega) \propto \alpha_{s}(\Omega) \delta p_{ref}
\end{equation}
With this detection scheme the phase and amplitude fluctuations can be retrieved. In order to measure both amplitude and phase fluctuations at the same time to study their correlations, they are measured by collecting the signal of each detector and numerically calculating the sum and difference.

\section{Normalisation covariance matrices:}
\label{supp:normalisation}
Once the data acquired, they need to be normalised before interpretation. The first step is to normalise the data with respect to the shot noise.\\
According to equation (\ref{eq.:delt_I_+}) and taking into account the dark noise as well as a pixel dependent gain, the measured intensity can be expressed as
\begin{equation}
\delta \mathcal{I}_{+,i} = g_i \delta x_{s,i} + d_i
\end{equation}
Where $g_i$ is a variable gain, $d_i$ the dark noise and $i$ is the pixel's index. Taking the covariance of the measured intensity, assuming no correlations between the signal and the dark noise, leads to
\begin{equation}
\mathrm{cov} \left[ \delta \mathcal{I}_{+} \right]_{i,j} = g_i g_j \mathrm{cov} \left[ \delta x_{s} \right]_{i,j} + \mathrm{cov} \left[ d \right]_{i,j}
\end{equation}
To determine the gain factor, the variance of the signal is calculated at a high frequency so that the only noise measured is the shot noise, i.e. $\mathrm{var} \left[ \delta x_{s} \right]_{i, shot} = 1$. Thus, $g_i$ is given by
\begin{equation}
\label{eq.:gain}
g_i = \sqrt{\mathrm{var} \left[ \delta \mathcal{I}_{+} \right]_{i, shot} - \mathrm{var} \left[d \right]_i }
\end{equation}
Finally, the amplitude quadrature is given by
\begin{equation}
\label{eq.:amp_quad_cov}
\mathrm{cov} \left[ \delta x_{s} \right]_{i,j} = \frac{\mathrm{cov} \left[ \delta \mathcal{I}_{+} \right]_{i,j} - \mathrm{cov} \left[ d \right]_{i,j}}{g_i g_j}
\end{equation}

\paragraph*{}
The normalisation is slightly more complicated for the phase quadrature. As seen in equation (\ref{eq.:delt_I_-}), the phase fluctuations measured, $\delta \mathcal{I}_{-,i} \propto \alpha_{s,i}\alpha_{ref,i}\delta \phi_{s,i}$, are proportional to the number of photons in the reference field, which is weaker than the signal ($\frac{\alpha_{s}^2}{\alpha_{ref}^2} \simeq 50$). Consequently, the phase fluctuations measured are not directly the fluctuations from the signal but the ones from the reference. It can be seen as the signal fluctuations measured after a loss channel. Those losses can easily be modelled by a beam splitter where one input is the signal and the other one the vacuum. This BS would have a reflectivity $r_i = \frac{\alpha_{ref,i}}{\alpha_{s,i}}$ and a transmission $t_i = \sqrt{1- \frac{\alpha_{ref,i}^2}{\alpha_{s,i}^2}}$. Thus, the measured phase fluctuations and signal ones are related by
\begin{equation}
\delta p_{ref,i} = \frac{\alpha_{ref,i}}{\alpha_{s,i}} \delta p_{s,i} + \sqrt{1- \frac{\alpha_{ref,i}^2}{\alpha_{s,i}^2}} \delta p_{v,i} = r_i \delta p_{s,i} + t_i \delta p_{v,i}
\end{equation}
Where $\delta p_v$ are the phase fluctuations of the vacuum.\\
As previously, the intensity fluctuations are related to the phase ones by 
\begin{equation}
\delta \mathcal{I}_{-,i} = g_{i}\delta p_{ref,i} + d_i
\end{equation}
Where here again, $g_i$ is a variable gain and $d_i$ the dark noise. They are not necessarily the same as for the amplitude fluctuations but the same notation is use for simplicity.\\
Using the previous expression for $\delta p_{ref}$, the intensity fluctuations can be written as
\begin{equation}
\delta \mathcal{I}_{-,i} = g_i r_{i} \delta p_{s,i} + g_i t_i \delta p_{v,i} + d_i
\end{equation}
As none of those contributions are correlated, the covariance is given by
\begin{eqnarray}
\mathrm{cov} \left[ \delta \mathcal{I}_{-} \right]_{i,j}  =  g_i g_j r_i r_j \mathrm{cov} \left[ \delta p_{s} \right]_{i,j} \\ \nonumber
\ + g_i g_j t_i t_j \mathrm{cov} \left[ \delta p_v \right]_{i,j}+ \mathrm{cov} \left[ d \right]_{i,j}
\end{eqnarray}
Once again, to determine the gain factor, the variance of the signal is measured at a high frequency so that var$\left[ \delta p_s \right]_{i,shot} = 1$. Furthermore, var$\left[ \delta p_v \right]_i = 1$, as the vacuum is uncorrelated. Finally, knowing that $r_{i}^2 + t_{i}^2 = 1$, the same equation as (\ref{eq.:gain}) is found for gain:
\begin{equation}
g_i = \sqrt{\mathrm{var} \left[ \delta \mathcal{I}_{-} \right]_{i, shot} - \mathrm{var} \left[d \right]_i }
\end{equation}
Eventually, the covariance matrix for the signal phase fluctuations is given according to
\begin{equation}
\mathrm{cov}\left[ \delta p_s \right]_{i,j} = \frac{\mathrm{cov} \left[ \delta \mathcal{I}_{-} \right]_{i,j} - \mathrm{cov} \left[ d \right]_{i,j}}{g_i g_j r_i r_j} - \frac{t_i t_j}{r_i r_j} Id
\end{equation}
Where $Id$ is the identity matrix. As a matter of fact, the covariance of the vacuum fluctuations is the identity matrix because no correlations exist between the different frequency ranges.

\end{document}